\newcommand\bb[1] {   \mbox{\boldmath{$#1$}}  }

\newcommand\del{\bb{\nabla}}
\newcommand\bcdot{\bb{\cdot}}

\newcommand\vv{\bb{v}}
\newcommand\B{\bb{B}}
\newcommand\BV{Brunt-V\"ais\"al\"a\ }

\newcommand\kva{ \bb{k\cdot v_A}  }

    \def\dd{\partial}
    \def\tilde{\widetilde}

    \def\beq{ \begin{equation} }
    \def\eeq{ \end{equation} }
    \def\spose#1{\hbox to 0pt{#1\hss}}
    \def\ltsim{\mathrel{\spose{\lower.5ex\hbox{$\mathchar"218$}}
	 \raise.4ex\hbox{$\mathchar"13C$}}}

\def\tilde{\widetilde}

\documentclass{mn2e}

\input psfig

\title[Differential Rotation in Weakly-Magnetized NS Atmospheres]
	{Differential Rotation in Weakly-Magnetized Neutron Star Atmospheres}
	
\author[K. Menou]
       {Kristen Menou$^{1}$\\
	$^1$Institut d'Astrophysique de Paris, 98bis Boulevard Arago, 
	75014 Paris, France}	

\begin{document}

\maketitle

\begin{abstract}

The atmospheres of weakly-magnetized neutron stars expand
hydrostatically and rotate differentially during thermonuclear X-ray
bursts.  Differential rotation is probably related to the frequency
drifts of millisecond burst oscillations exhibited by about a dozen
nuclear-powered X-ray pulsars.  Here, we analyze the linear stability
of this differential rotation with respect to local, axisymmetric,
multi-diffusive MHD perturbations, at various heights in the neutron
star atmosphere.  Unstable magneto-rotational modes are identified
from within to well above the burning layers.  Properties of the
fastest growing modes depend sensitively on the local magnetic field
geometry. Linear estimates suggest that momentum transport due to
magneto-rotational instabilities can affect atmospheric rotation
profiles on time-scales relevant to burst oscillations. This transport
would likely strengthen the coherence of burst oscillations and
contribute to their drifts.

\end{abstract}


\section{Introduction}

Since their discovery in the 1970s (see Lewin, van Paradijs \& Taam
1995 for a review), thermonuclear X-ray bursts have been observed from
a large number of low-mass X-ray binaries containing weakly-magnetized
accreting neutron stars. The basic physical mechanism responsible for
X-ray bursts, namely mass accumulation on the neutron star surface at
rates such that a thermonuclear He or mixed H/He flash is eventually
triggered, is well understood (see, e.g., Bildsten 1998 and references
therein). The phenomenology of X-ray bursts is rich, however, and
theoretical models face some difficulties when detailed comparisons
with the observations are attempted (e.g. Bildsten 2000; Galloway et
al. 2003).

This situation has become more serious, and more exciting, in the past
decade, with the discovery of transient millisecond ($300$--$600$~Hz)
oscillations during some X-ray bursts, the so-called burst
oscillations (Strohmayer et al. 1996). While the presence of
oscillations during burst rise may be understood as resulting from the
rotational modulation of a nuclear burning spot spreading around the
stellar surface (Spitkovsky, Levin \& Ushomirsky 2002), it is unclear
why oscillations should be visible past the burst peak, when the
burning fuel has presumably been entirely ignited. Also intriguing is
the nature of frequency drifts exhibited by most burst oscillations
(generally spin-ups, by a few Hz at most) and the origin of the
diversity they show in terms of drift time-scales and amplitudes (see
Strohmayer \& Bildsten 2003 for a review).  Observations of the burst
oscillation phenomenon have been very rewarding. In particular, it has
recently been shown that the burst oscillation asymptotic frequency
(after complete spin-up) and the independently-known neutron star spin
frequency are closely related in two millisecond accreting pulsars
(Chakrabarty et al. 2003; Strohmayer et al. 2003). Still, our
understanding of the physical mechanisms involved in burst
oscillations and associated frequency drifts remains limited.

Following Strohmayer et al. (1997), Cumming \& Bildsten (2000)
proposed that frequency drifts originate from the rotational evolution
of heated burning layers, as they cool, hydrostatically contract and
nearly recover their original rotation, assuming their specific
angular momentum is conserved at all times. These same authors later
realized, however, that the largest observed frequency drifts are in
excess of the amplitudes expected in their burning layer scenario, by
as much as a factor 2-3. Apparently, only layers higher up in the
atmosphere would have the right amount of rotational offset to explain
the largest drift amplitudes (Cumming et al. 2002). More recently,
Chakrabarty et al. (2003) have also argued that the very rapid drift
(spin-up) that they have observed during the rise phase of one of the
bursts of SAX J1808-3658 effectively rules out the scenario of Cumming
\& Bildsten (2000). There is currently no solid alternative
explanation for the origin of burst oscillations and associated
frequency drifts, even though both Rossby wave (Heyl 2003; but see Lee
2003) and zonal flow (Spitkovsky et al. 2002) scenarios have been
proposed.

In their study of neutron star atmospheres, Cumming \& Bildsten (2000)
have analyzed a number of mechanisms which could make the vertical
differential rotation established via burst-induced hydrostatic
expansion unstable. They have not identified any clear destabilizing
process (which would presumably lead to angular momentum transport
between atmospheric shells) and this was the main motivation for their
assumption of conserved specific angular momentum.  The purpose of the
present study is to reconsider the issue of stability of differential
rotation in the atmospheres of weakly-magnetized, rapidly-rotating
neutron stars.

As noted by Cumming \& Bildsten (2000), the strong vertical thermal
stratification of a neutron star atmosphere stabilizes it with respect
to adiabatic perturbations, even in the presence of strong vertical
differential rotation. Although not considered by these authors in
their analysis, one indeed verifies that the vertical differential
rotation expected in the present context is stable according to both
the hydrodynamical Solberg-H\o iland adiabatic criteria (e.g. Tassoul
1978) and their generalization for weakly-magnetized fluids (Balbus
1995).

This situation is reminiscent of the solar radiative interior, where
thermal stratification also plays a strongly stabilizing
role. Goldreich \& Schubert (1967) and Fricke (1968) have shown,
however, that rotational instabilities can still exist, provided
perturbed fluid elements are allowed to exchange heat with their
environment much faster than they exchange momentum. The stabilizing
role of the atmospheric thermal stratification is then effectively
neutralized when a perturbed fluid element reaches thermal equilibrium
with its environment while its original momentum remains largely
unchanged. In the solar interior, radiative heat diffusion is indeed
orders of magnitude faster than viscous diffusion of momentum, and
this was the initial motivation for the double-diffusive analysis of
Goldreich \& Schubert and Fricke (hereafter GSF altogether). The basic
double-diffusive mechanism invoked in these instabilities is not too
different from that of ``salt-finger'' instabilities operating in the
Earth's oceans, except that the destabilizing salinity stratification
is replaced by a destabilizing angular momentum stratification in the
stellar context.

As we shall see below, conditions in the atmospheres of
weakly-magnetized neutron stars during bursts are also strongly
multi-diffusive. It is then possible that rotational instabilities
will develop and affect the differentially-rotating atmospheric layers
in such a way as to influence the observational properties of burst
oscillations. To address this question properly, it is necessary to
account for the presence of magnetic fields in neutron star
atmospheres. Recently, Menou, Balbus \& Spruit (2004) have generalized
the work of GSF to weakly-magnetized fluids. They have shown that the
combination of weak magnetic fields and a multi-diffusive situation
may have contributed to establishing the state of near solid-body
rotation inferred from helioseismology for the solar interior.

In this paper, we present a specific application of the stability
analysis of Menou et al. (2004) to weakly-magnetized,
differentially-rotating neutron star atmospheres and we discuss
possible consequences for the phenomenology of burst oscillations and
associated frequency drifts. We describe our method of solution in
\S2. In \S3, we present the results of our stability analysis at
various heights in the neutron star atmosphere. The relevance of these
results to burst oscillations is discussed in \S4.

\section{Method}

Following Menou et al. (2004), we perform a local, linear stability
analysis of differential rotation in a weakly-magnetized,
stably-stratified fluid. In this section, we first recall the
dispersion relation obeyed by axisymmetric modes in a general context
in \S2.1. We describe how this dispersion relation can be applied to
the specific problem of weakly-magnetized neutron star atmospheres
during bursts in \S2.2. We then describe how we obtained numerical
solutions to the dispersion relation in \S2.3. In \S2.4, we discuss
the issue of the efficiency of momentum transport, that we estimate
from linear mode properties.  Our basic analysis follows closely that
presented in Menou et al. (2004), where additional details on the
method can be found.

\subsection{Dispersion Relation}

We initially work in cylindrical coordinates, $({R}, {\phi}, {Z})$,
but we will also later express results in terms of spherical
coordinates, $(r, \theta, \phi)$. We consider axisymmetric Eulerian
perturbations with a WKB space-time dependence $\exp [i(\bb{k\cdot r}
- \omega t)]$, where $\bb{k}= (k_R, 0, k_Z) = (k_r,k_\theta,0)$.
Starting from the set of MHD equations (continuity, momentum,
induction, energy) that include the effects of viscosity, resistivity
and heat conduction

\begin{eqnarray}
& & {\dd\rho\over \dd t} + \del\bcdot (\rho \bb {v}) =  0,
\end{eqnarray}
\begin{eqnarray}\label{mom}
& & \rho {\dd\vv \over \dd t} + (\rho \vv\bcdot\del)\vv = -\del\left(
P + {B^2\over 8 \pi} \right)  - \rho \del \Phi \nonumber \\
& & + \left( {\B\over 4\pi}\bcdot \del\right)\B + \mu \left( \del^2 \vv + 
\frac{1}{3} \del(\del \bcdot \vv) \right),
\end{eqnarray}
\begin{eqnarray}
& & {\dd\bb{B}\over \dd t} = \bb {\nabla\times (v\times B)} - 
  \eta { \bb \nabla\times} \left( {\bb \nabla\times \bb{B}} \right),
\end{eqnarray}
\begin{eqnarray}
& & {1\over (\gamma -1)} {P} {d\ln P\rho^{-\gamma}\over dt} = \chi \del^2 T,
\end{eqnarray}

one derives,  under the Boussinesq approximation and ignoring self-gravity, the following fifth-order dispersion relation for local,
axisymmetric, multi-diffusive modes, 

\begin{eqnarray}  
\nonumber & & {\tilde\omega_{b+v}}^4 {\omega_{e}}\, {k^2\over k_Z^2} + {\tilde\omega_{b+v}}^2 {\omega_{b}}
\left[ {1\over \gamma \rho}\, \left({\cal D} P\right)\, {\cal D} \ln
P\rho^{-\gamma}\right]\\ 
& & +{\tilde\omega_{b}}^2 {\omega_{e}} \left[
{1\over R^3}\, {\cal D} (R^4\Omega^2) \right] - 4 \Omega^2 (\kva)^2 {\omega_{e}}=
0,\label{eq:disprel}
\end{eqnarray}

where
\begin{eqnarray}
& &\bb{v_A} = { \bb{B}/\sqrt{4\pi\rho}}, \qquad 
k^2 = k_R^2 + k_Z^2,\nonumber \\ 
&& {\tilde\omega_{b+v}}^2=\omega_b \omega_v -(\kva)^2,
\qquad {\tilde\omega_{b}}^2=\omega_b^2 -(\kva)^2, \nonumber\\
\nonumber & & \omega_b=\omega+i \eta k^2, \qquad  \omega_v=\omega+i \nu k^2,\\
 \qquad
& & \omega_e=\omega+ \frac{\gamma -1}{\gamma}{i  T\over  P} \chi k^2, 
\qquad {\cal D} \equiv \left( \frac{k_R}{k_Z}\frac{\dd}{\dd Z}
-\frac{\dd}{\dd R}\right). \nonumber
\end{eqnarray}

A developed form of the dispersion relation can be found in Menou et
al. (2004). The \BV frequency, $N$, which measures the magnitude of
atmospheric thermal stratification, comes out of the first bracket
term above. Similarly, the epicyclic frequency, which measures the
angular momentum ``stratification,'' comes out of the second bracket
term. The notation is standard and identical to that adopted in Menou
et al. (2004): $\bb{v}$ is the fluid velocity, $\rho$ is the mass
density, $P$ is the pressure, $\Phi$ is the gravitational potential,
$T$ is the temperature, $\bb{B}$ is the magnetic field, $\mu$ is the
dynamic viscosity, $\nu = \mu / \rho$ is the kinematic viscosity,
$\eta$ is the resistivity, $\chi$ is the heat conductivity and
$\bb{v_A}$ is the Alfv\'en speed.

A value $\gamma = 5/3$ is adopted for the gas adiabatic index and a
perfect gas equation of state is assumed. Models of neutron star
atmospheres during thermonuclear X-ray bursts show that radiation
pressure sometimes contribute a non-negligible fraction to the total
pressure, especially close to the burning layers (see, e.g., Cumming
\& Bildsten 2000).  We neglect this contribution here, but it should
be included for better accuracy. {Degeneracy is neglected, because it
is lifted during bursts, and general relativistic effects are ignored.
When discussing burning layers, we also ignore the effects that any
remaining nuclear burning may have on stability, both as a heat source
and in modifying the composition of burning regions.}  The basic state
rotation is given by $\bb{\Omega} = (0, 0, \Omega(R,Z))$ along the
${Z}$--axis. The basic state magnetic field, whose geometry is
specified below, is assumed to be weak with respect to both rotation
and thermal pressure.

\subsection{Neutron Star Atmosphere During Burst}

\begin{table*}
{\footnotesize
\begin{center}
\begin{tabular}{cccccccccc}
\hline
\\
Layer & $\rho$&$T$&$\kappa$&$\nu$&$\nu_r$&$\eta$&$\xi_{rad}$&$\epsilon_{\nu}$&$\epsilon_{\eta}$\\
&(g cm$^{-3}$)&($10^8$ K)&(cm$^2$ g$^{-1}$)&(cm$^2$ s$^{-1}$)&(cm$^2$ s$^{-1}$)&(cm$^2$ s$^{-1}$)&(cm$^2$ s$^{-1}$)&$(\nu_{\rm tot} / \xi_{rad})$ & $(\eta / \xi_{rad})$\\
\\
\hline
\\
Burning & $10^5$& $10$ & $0.15$&$88$&$34$&$0.13$&$4.5 \times 10^5$&$2.7 \times 10^{-4}$&$2.9 \times 10^{-7}$\\
Middle & $3 \times 10^3$& $5$ & $0.17$&$680$&$2.5 \times 10^3$&$0.28$&$5.6  \times 10^7$&$5.7 \times 10^{-5}$&$5 \times 10^{-9}$\\
High & $100$ & $2$ & $0.23$&$2.1 \times 10^3$ &$4.4 \times 10^4$&$1.1$&$2.4 \times 10^9$ & $1.9 \times 10^{-5}$ & $4.6 \times 10^{-10}$\\
\\
\hline
\end{tabular}
\caption{Typical Atmospheric Conditions During Burst}
\end{center}}
\label{tab:one}
\end{table*}

In order to solve the above dispersion relation
(Eq.~[\ref{eq:disprel}]), we must specify values for all the physical
parameters entering the problem. Often, we rely on the recent,
one-dimensional, hydrostatic models of Cumming \& Bildsten (2000) for
neutron star atmospheres during bursts. We consider conditions in
three representative layers, right after burst trigger and the
hydrostatic expansion that follows it. The first such layer, labeled
"Burning" in Table~1, is representative of conditions in the burning
layers. The second layer, labeled "High", corresponds to typical
conditions high in the atmosphere, well above the burning layers but
still well below the stellar photosphere (conditions provided by
A. Cumming, private communication).  The third layer, labeled
"Middle", corresponds to atmospheric conditions intermediate between
the first two. Our choice of layers, although somewhat arbitrary, will
prove to be useful to illustrate variations of the stability
properties with height in the neutron star atmosphere.  We list in
Table~1 the values of the density, $\rho$, temperature, $T$, and
Rosseland-mean radiative opacity, $\kappa$ (from electron scattering),
in the three layers of interest.

We need to estimate the values of various microscopic diffusion
coefficients relevant to the stability problem.  From Spitzer (1962)
and Schwarzschild (1958), assuming that the plasma is mostly composed
of hydrogen, the (ion-dominated) dynamic viscosity, $\mu$, the
radiative kinematic viscosity, $\nu_r$, the electric resistivity,
$\eta$, the radiative heat conductivity, $\chi_{rad}$, and the
corresponding heat diffusivity, $\xi_{rad}$, are given by

\begin{eqnarray}
\mu &=&\rho \nu \simeq 2.2 \times 10^{-15} \frac{T^{5/2}}{\ln \Lambda}~
{\rm g~cm^{-1}~s^{-1}}, \\
\nu_r &=& \frac{16}{15} \frac{\sigma T^4} {\kappa \rho^2 c^2}~{\rm cm^2~s^{-1}}, \\
\eta &\simeq& 5.2 \times 10^{11} \frac{\ln \Lambda}{T^{3/2}}~
{\rm cm^2~s^{-1}}, \\
\chi_{rad} &=& {16T^3\sigma\over 3 \kappa\rho}~{\rm c.g.s.}, \\
\xi_{rad} &=& \frac{\gamma-1}{\gamma} \frac{T}{P} \chi_{rad}~{\rm cm^2~s^{-1}},
\end{eqnarray}

where $\ln \Lambda $ is the Coulomb logarithm, $c$ is the speed of
light and $\sigma$ is the Stefan-Boltzmann constant.  We list in
Table~1 the values of all three diffusivity coefficients for the three
layers of interest.  In addition, the values of the Prandtl number,
$\epsilon_\nu = \nu_{\rm tot} / \xi_{rad}= (\nu + \nu_r) / \xi_{rad}$,
and the ``Acheson number,'' $\epsilon_\eta = \eta / \xi_{rad}$, which
together measure the hierarchy of diffusive processes, are listed.
Viscous diffusion clearly dominates over resistive diffusion in the
neutron star atmosphere (a situation which is opposite to that in the
Sun's radiative interior; Menou et al. 2004), but radiative heat
diffusion is the strongest by many orders of magnitude (as is the case
for the solar interior). It is this property of rapid heat diffusion
that allows multi-diffusive modes to overcome the stabilizing
influence of the atmospheric thermal stratification. Note also how the
situation becomes more double-diffusive as one goes up in the
atmosphere (since the values of $\epsilon_\nu$ and $\epsilon_\eta$
decrease), and how radiative viscosity eventually dominates over ion
viscosity high enough above the burning layers.

A number of global parameters for the magnetized,
differentially-rotating neutron star atmosphere must be specified. We
adopt a value $\Omega =2 \times 10^{3}$~s$^{-1}$ for the neutron star
angular velocity, corresponding to a typical spin period of $\sim
3$~ms (spin frequency $\sim 300$~Hz).  We adopt a value $N/\Omega
\simeq 10^2$ for the ratio of the \BV frequency to the angular
velocity (Cumming \& Bildsten 2000). Note that we only include the
thermal stratification in our definition of the \BV frequency since
the dispersion relation that we are solving was derived ignoring the
effects of composition gradients. The issue of stabilization due to
composition gradients is important and is further discussed in \S4.

We need to specify the level of differential rotation present in the
atmosphere. After burst trigger, all the atmospheric layers, from the
burning regions to those located higher up, are heated up. This occurs
on a thermal timescale, which is much in excess of a local sound
crossing time (Cumming \& Bildsten 2000) and, as a result, the layers
expand hydrostatically. Assuming conservation of specific angular
momentum during this phase, differential rotation between spherical
shells should then be established at a level corresponding to $\dd \ln
\Omega / \dd \ln r =-2$, in all three layers of interest. We further
assume that no differential rotation is present within spherical
shells. It should be noted that any such negative differential
rotation (i.e. $\dd \ln \Omega / \dd \theta < 0$) would be unstable to
adiabatic magneto-rotational modes (Balbus 1995; Menou et al. 2004).

Finally, we must specify a geometry and a strength for the basic state
magnetic field present in the atmospheric layers. Although other
configurations are possible (for instance with high multipoles; see
also Ruderman 2003), we will assume here, for simplicity, a misaligned dipole
geometry. More precisely, because our analysis is axisymmetric in
nature, we must  use an "axisymmetric dipole"
geometry, which best resembles that of a true dipole.   The
spherical-radial and polar components of the field are thus

\beq
B_r=B_p \cos (\theta-\theta_{\rm mis}), ~~B_\theta=\frac{B_p}{2} \sin
(\theta-\theta_{\rm mis}),
\eeq

where $\theta$ is the polar angle ($\theta=0$ or $\pi$ on the neutron
star rotation axis), $\theta_{\rm mis}$ is the equivalent of the angle
by which the dipolar magnetic axis is misaligned from the rotation
axis and $B_p$ is the polar field strength at the neutron star surface
(assumed to be the same in all layers of interest). With this
geometry, field lines have an axisymmetric configuration such that
they are nearly vertical for angles $\theta \sim \theta_{\rm mis}$
(magnetic "pole," which is an axisymmetric band, really) and nearly
horizontal at the magnetic equator.  This setup allows us to study the
effects of the magnetic field geometry on the stability properties. We
have investigated stability in the three atmospheric layers of
interest for values of $B_p$ ranging from $10^7$ to $10^9$~G,
considered as typical for weakly-magnetized accreting neutron stars.

\subsection{Numerical solutions}

The complexity of the dispersion relation (Eq.~[\ref{eq:disprel}])
makes it difficult to analytically derive necessary and sufficient
conditions for stability (Menou et al. 2004). In addition, since we
are interested in studying the properties of unstable modes, we want
to solve the dispersion relation numerically, with the appropriate set
of parameters.  To do this, we use the Laguerre algorithm described by
Press et al. (1992) to solve equation~(\ref{eq:disprel}) as a
fifth-order complex polynomial for $\sigma=-i \omega$, with real
coefficients.

We have found useful to rewrite both the rotational and thermal
stratification bracket terms appearing in the dispersion relation in
spherical coordinates $(r, \theta, \phi)$. The thermal stratification
term then becomes function only of $N^2$, $\theta$ and the radial and
angular wavevectors, $k_r$ and $k_\theta$, for a spherically symmetric
star.  The rotational stratification term, on the other hand,
explicitly depends on the amount of differential rotation within and
between spherical shells, that we express as $\dd \ln \Omega / \dd
\theta$ and $\dd \ln \Omega / \dd \ln r$, respectively. As we have
mentioned earlier, we are interested here in the specific case
corresponding to $\dd \ln \Omega / \dd \theta = 0$ and $\dd \ln \Omega
/ \dd \ln r = -2$.

For each of the three atmospheric layers of interest, we search for
unstable modes (which correspond to $\sigma$-roots with positive real
parts) at various locations on the stellar sphere.  We consider values
of the polar angle, $\theta$, in the range $0 \to \pi/2$ (pole to
equator). At each location, we perform a search for unstable modes by
varying the wavevectors $k_r$ and $k_\theta$ independently in the
range $\pm [2 \pi / l_{\rm max}, 2 \pi / l_{\rm min} ]$. The first,
large-scale limit, $l_{\rm max}$, is chosen to be $\ltsim H$, the
pressure scale height, so that the local and weak field assumptions of
our analysis remain valid. The second, small-scale limit, $l_{\rm
min}$, is chosen to be $ \gg \lambda$, the mean free path, so that the
analysis remains valid given our use of fluid equations. We estimate
$H \simeq \Re T/ g$, where $\Re$ is the perfect gas constant, $\lambda
\sim \nu / \sqrt{\Re T}$ and we use $g \simeq 2 \times
10^{14}$~cm~s$^{-2}$ as a typical neutron star surface gravity.  In
addition, we systematically perform focused searches in regions of the
wavevector space corresponding to nearly cylindrical-radial
displacements ($|k_R/k_Z| \to 0$) and nearly vertical displacements
($|k_R/k_Z| \to \infty$).

\subsection{Efficiency of Transport}

The differential rotation established in neutron star atmospheres
during bursts is an intrinsically transient phenomenon. Absent any
transport mechanism, the various differentially-rotating spherical
shells would essentially recover their original rotation (modulo a
small offset due to changes in molecular weight; Cumming \& Bildsten
2000) once the atmosphere cools down to pre-burst conditions,
typically after several tens of seconds. In the presence of a
transport mechanism, however, the issue of the efficiency of momentum
transport is of direct relevance to the observations because a
mechanism effective enough to act on time-scales of tens of seconds or
shorter would influence the rotational evolution of atmospheric layers
during burst.

It is unclear whether one can determine, from a linear analysis only,
the efficiency of transport that would result from non-linear growth
and turbulence. We will adopt an often-used prescription here, by
assuming that the efficiency of transport can be crudely estimated
from the growth rate and length-scale of the fastest-growing unstable
linear mode. There is no general justification for this simple
prescription, except perhaps examples of physical situations for which
it is not too far off (for instance, the emergence of mid-latitude
weather structures in the Earth's turbulent atmosphere with sizes
matching linear predictions for the fastest growing baroclinic modes
is considered a success of baroclinic theory; see Pedlosky 1987). Our
goal here is not to accurately estimate the efficiency of momentum
transport between shells, but merely to question the observational
relevance of this transport given typical burst durations.

It should be noted that the non-linear numerical simulations of
Korycansky (1991) lend support to the simple prescription we
adopted. This author found, for an equatorial setup (i.e. aligned
thermal and angular momentum stratifications), that the non-linear
behavior of hydrodynamical double-diffusive (GSF) modes is well
described by a combination of linear growth rate and spherical-radial
length scale of the instability. Assuming this property carries over
to magnetized modes and to other geometries (i.e. away from the
stellar equator), we will be estimating the efficiency of momentum
transport across spherical shells with the quantity $\sigma_{\rm
max}/(k_r^2)_{\rm max}$ (dimension of a viscosity), where $\sigma_{\rm
max}$ is the growth rate of the fastest-growing mode and $(k_r)_{\rm
max}$ is the spherical-radial wavevector component of that same
mode. Clearly, using linear theory to predict a non-linear outcome is
a daring extrapolation, so one should treat our results on transport
efficiencies with much caution.

\section{Results}

\begin{table*}
{\footnotesize
\begin{center}
\begin{tabular}{cccccc}
\hline
\\
Model &Layer &$\frac{\partial \ln \Omega}{\partial \ln r}$&$B_p$&$\theta_{\rm mis}$&Special \\
& & &(Gauss)&(degrees)&\\
\\
\hline
\\
B1 & Burning & $-2$ &$10^8$ & $0$ & \\
B2 & Burning & $-2$ &$10^7$ & $0$ & \\
B3 & Burning & $-2$ &$10^8$ & $30$ & \\
\\
M1 & Middle & $-2$ &$10^8$ & $30$ & \\
M2 & Middle & $-2$ &$10^8$ & $0$ & \\
M3 & Middle & $-2$ &$10^8$ & $20$ & \\
M4 & Middle & $-2$ &$10^8$ & $30$ & $\Omega \times 2$\\
\\
H1 & High & $-2$ &$10^7$ & $30$ & \\
H2 & High & $-1$ &$10^7$ & $30$ & \\
H3 & High & $-2$ &$10^7$ & $30$ & large $l_{\rm max} \sim R_{\rm NS}$\\

\\
\hline
\end{tabular}
\caption{Models and Parameters}
\end{center}}
\label{tab:two}
\end{table*}

For each of the three atmospheric layers of interest, we have run
several models to identify parameter dependencies. To best illustrate
these dependencies, we have chosen a specific subset of models, listed
in Table~2.  In most cases, only the magnetic field strength, $B_p$,
or the value of $\theta_{\rm mis}$, the angle by which the magnetic
``pole'' is misaligned from the rotation axis, are varied. But we also
show the effects of variations in the neutron star rotation rate,
$\Omega$, the rate of spherical-radial differential rotation,
$\partial \ln \Omega / \partial \ln r$, and the maximum allowed scale
for perturbations.

Quite generally, many unstable modes were found in the searched
wavevector space, as expected for the strongly multi-diffusive
conditions considered.  {All the unstable modes that we have
identified were of the ``direct'' type (i.e. with $Im(\sigma)=0$), as
opposed to the other multi-diffusive ``overstable'' type (when
$|Im(\sigma)| > Re(\sigma) >0$). This is a consequence of the strongly
stabilizing thermal stratification and of the fast rate of heat
diffusion (relative to momentum diffusion), as was the case for the
solar radiative zone (Menou et al. 2004).} To simplify the description
of these unstable modes, we focus our attention on the properties of
the fastest-growing mode identified, as a function of location on the
stellar sphere (i.e. polar angle, $\theta$).  While we cannot
guarantee that we have identified the single fastest-growing mode at
any location, convergence tests with increasingly higher resolutions
for the wavevector grid and the polar angle grid were successfully
passed, suggesting that our search method is robust. It is worth
pointing out that we have often experienced unsatisfactory numerical
convergence for small values of the polar angle, i.e. $\theta \sim 0$
(close to the rotation axis). This may be caused by the condition for
a weak magnetic field compared to rotation becoming increasingly more
stringent at small $\theta$ values (where pressure gradients become
nearly orthogonal to the centrifugal force), but we have not explored
this possibility further. We also do not explicitly show results for a
magnetic field strength $B_p = 10^9$~G. For reasons that will become
clear by the end of this section, searches for unstable modes in this
stronger field situation have been unsuccessful high in the
atmosphere.

We begin with a discussion of the general properties of
fastest-growing unstable modes identified in the burning layers,
before comparing them with the properties of fastest-growing modes
identified higher-up in the atmosphere.

\subsection{Burning Layers} 

\begin{figure}
\psfig{figure=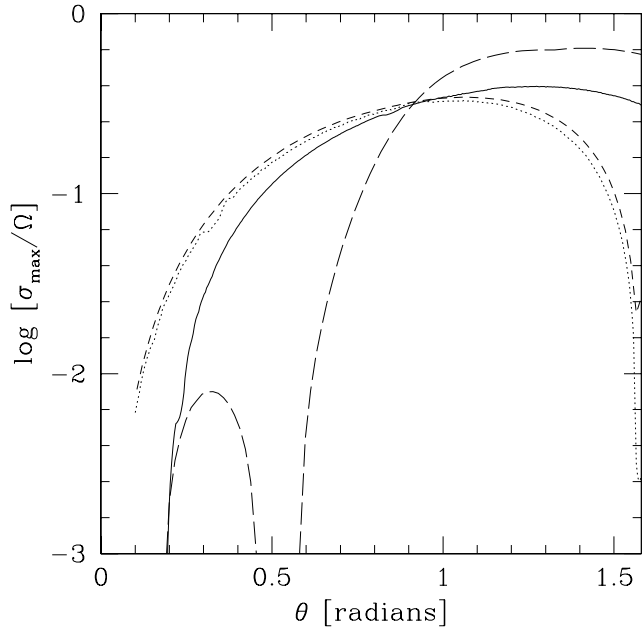,width=3.5truein,height=3.5truein}
\caption{Growth rate of the fastest-growing mode, $\sigma_{\rm max}$,
in units of angular velocity, $\Omega$, as a function of polar angle,
$\theta$ (rotational equator at $\theta= \pi/2$). The solid, dotted,
short-dashed and long-dashed lines correspond to models B3, B1, B2 and
H1 in Table~2, respectively.}
\label{fig:one}
\end{figure}

Figure~\ref{fig:one} shows the maximum growth rate among all unstable
modes, in units of the neutron star angular velocity, $\Omega$, as a
function of polar angle, $\theta$.  The solid, dotted and short-dashed
lines correspond to models B3, B1 and B2 in Table~2,
respectively. These models differ in the strength of the magnetic
field and by how much the magnetic ``pole'' is misaligned from the
rotation axis at $\theta = 0$.

\begin{figure}
\psfig{figure=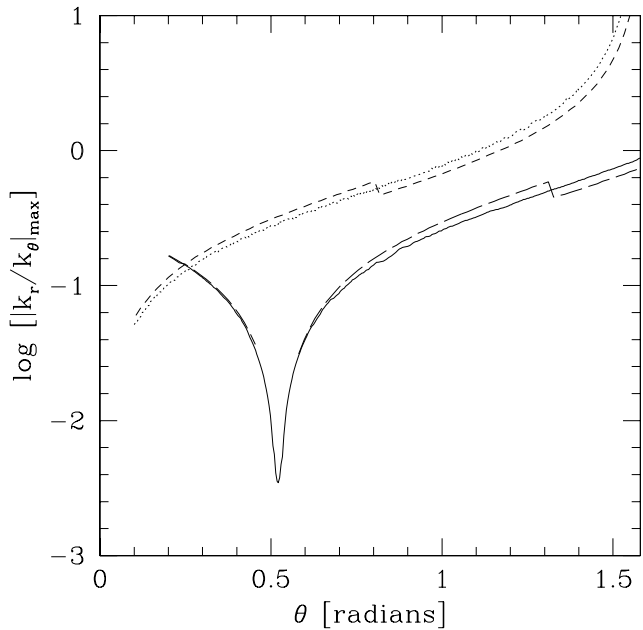,width=3.5truein,height=3.5truein}
\caption{Ratio of spherical-radial to polar wavenumber, $|k_r /
k_\theta|$, for the fastest-growing mode, as a function of polar angle,
$\theta$. Notation is identical to Fig.~1, for the same four models
(solid: B3; dotted: B1; short-dashed: B2; long-dashed: H1).}
\label{fig:two}
\end{figure}

In all three models, the growth rate of the fastest-growing mode is a
function of polar angle, $\theta$, which shows that the combination of
magnetic field and stellar spherical geometries have an influence on
stability properties.  The similarity between the dotted and
short-dashed lines also indicates that the magnetic field strength
does not strongly affect growth rates, although we will see later that
it does matter in a global sense (\S 3.2). On the other hand, there is
a clear distinction between the solid line (model B3) and the dotted
and short-dashed lines, indicating that the magnetic field geometry
alone influences the stability properties quite a lot.

These trends are confirmed by looking at the wavenumber properties of
the same modes. Figure~\ref{fig:two} shows the ratio of
spherical-radial to polar wavenumber, $|k_r/k_\theta|$, for the
fastest-growing modes, as a function of polar angle, $\theta$. The
notation is identical to Figure~\ref{fig:one}, for ease of
comparison. Once again, the magnetic field strength is found to have
little influence on the mode properties, but the magnetic field and
stellar spherical geometries do.\footnote{We have checked that the
discontinuities present in the short- and long-dashed lines in
figure~\ref{fig:two}, and absent from the corresponding growth rates
in figure~\ref{fig:one}, persist even at higher wavenumber or polar
angle grid resolutions. Consequently, they appear to have a physical
origin (possibly a change in the nature of the fastest-growing mode
with local geometry). We have not investigated this further.}

Figure~\ref{fig:two} reveals an interesting wavenumber property of
fastest-growing modes.  In the two models with a magnetic dipole
aligned with the rotation axis, the nature of the modes changes from
mostly spherical-radial ($|k_r/k_\theta| \ll 1$) for small $\theta$
values to mostly horizontal ($|k_r/k_\theta| \gg 1$) as one approaches
the stellar equator ($\theta \to \pi/2$). In the model with a magnetic
``pole'' misaligned by $\theta_{\rm mis} =30$ deg, on the other hand ,
$|k_r/k_\theta|$ is $\ll 1$ on a fair fraction of the stellar sphere,
although it approaches unity at the stellar equator. This behavior can
be understood as resulting mostly from the magnetic field geometry. At
the magnetic ``poles,'' where the field is nearly spherical-radial,
the fastest-growing mode tends to be nearly spherical-radial
itself,\footnote{The nearly incompressible modes described by our
dispersion relation are characterized by displacements orthogonal to
the mode wavevector, ${\bb k}$.} with $|k_r/k_\theta| \ll 1$. As one
approaches the magnetic equator, however, the field becomes nearly
horizontal and so does the fastest-growing mode. This is consistent
with a tendency for the fastest-growing mode to minimize the amount of
magnetic tension it is subject to, by favoring displacements nearly
along field lines.

\begin{figure}
\psfig{figure=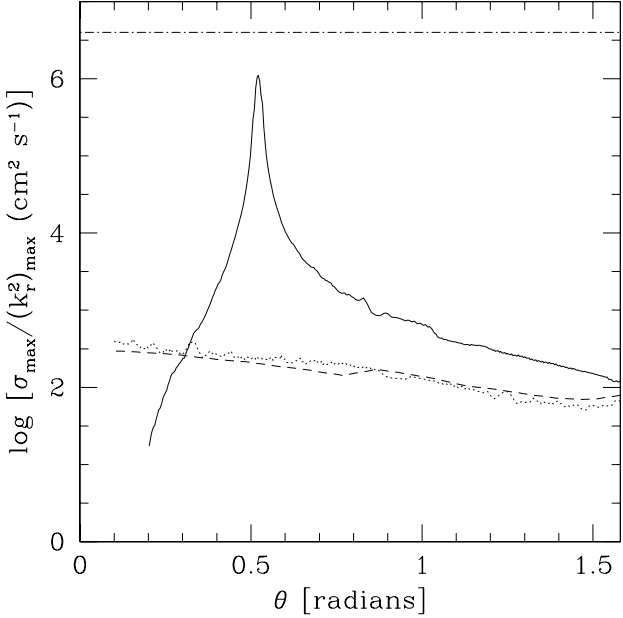,width=3.5truein,height=3.5truein}
\caption{Estimate of the efficiency of spherical-radial momentum
transport, $\sigma_{\rm max}/k_r^2$ (dimension of a viscosity), based
on fastest-growing mode properties in typical burning layer
conditions, as a function of polar angle, $\theta$. The solid, dotted
and dashed lines correspond to models B3, B1 and B2 in Table~2,
respectively.  The location of the magnetic ``pole'' at $\theta
=30$~deg in model B3 is obvious.  The horizontal dash-dotted line
indicates the efficiency required to affect the rotation profile over
a local scale-height in $1$~s.}
\label{fig:three}
\end{figure}

Figure~\ref{fig:three} shows estimates of the efficiency of momentum
transport in the spherical-radial direction, based on fastest-growing
linear mode properties, as a function of polar angle $\theta$. Again,
notation is similar to that in Figs.~1 and~2 for models B1 (dotted
line), B2 (dashed line) and B3 (solid line). In model B3, which
corresponds to a misaligned magnetic dipole, the estimated efficiency
of transport is strongly peaked at the magnetic ``pole'' (because of a
significant increase of the mode radial wavelength, as we shall see
below and in Fig.~\ref{fig:six}).  The horizontal dash-dotted line in
Figure~\ref{fig:three} indicates the efficiency typically required in
the burning layers to affect the rotation profile over a local
pressure scale-height, $H \simeq \Re T/ g$, in 1 second.  This
reference efficiency was simply calculated as
$H^2$~cm$^2$~s$^{-1}$. It would not make much sense to calculate a
reference efficiency for a lengthscale exceeding $H$ because both
atmospheric and stability conditions start to change significantly
over a local pressure scale-height.

Interestingly, the maximum momentum transport efficiency estimated
from fastest-growing mode properties, at the magnetic ``pole,'' falls
short of the reference efficiency only by a factor of a few,
suggesting that several seconds may be sufficient for the rotational
profiles to be affected. Away from the magnetic ``pole'' (and
essentially everywhere on the stellar sphere in aligned-dipole models
B1 and B2), estimated transport efficiencies appear too small, by
several orders of magnitude, to affect rotation profiles over a local
scale-height on observationally relevant time-scales. In fact, the
efficiencies indicated by the dashed and dotted lines in
Figure~\ref{fig:three} are small enough to be comparable to that of
microscopic viscosity ($\nu + \nu_r$) in the burning layers, as can be
seen from Table~2.

Many of the basic properties we have just described for
fastest-growing modes in burning layers carry over to higher
atmospheric layers. There are important differences, however, that we
will now illustrate. Rather than going through a description of growth
rates and wavenumbers independently for all cases of interest, we will
often summarize results by focusing on transport efficiencies (which
are a combination of the two) and how these are affected by changes in
physical conditions higher up in the atmosphere.

\subsection{High Atmosphere} 
 
\begin{figure}
\psfig{figure=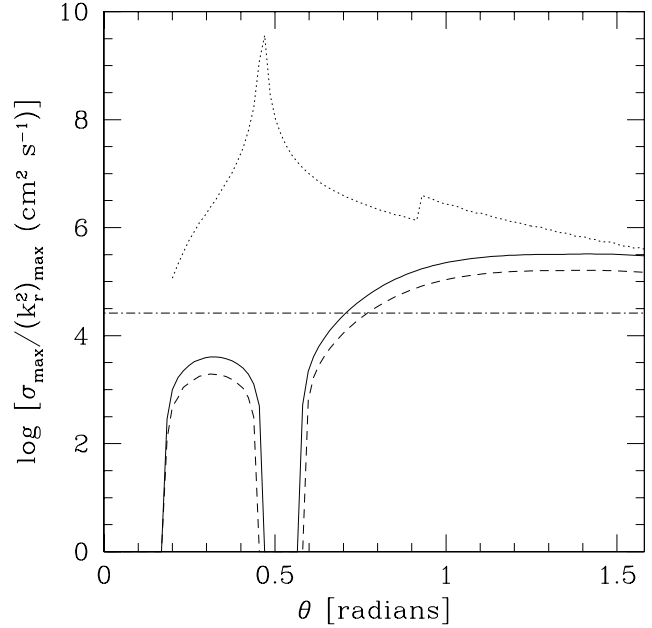,width=3.5truein,height=3.5truein}
\caption{Estimate of the efficiency of spherical-radial momentum
transport for typical conditions in a high-atmospheric layer, as a
function of polar angle, $\theta$. The solid, dotted and dashed lines
correspond to models H1, H3 and H2 in Table~2, respectively. The
horizontal dash-dotted line indicates the efficiency required to
affect the rotation profile over a local scale-height in $1$~s.}
\label{fig:four}
\end{figure}

In addition to the three burning layer models we have already
discussed, figures~\ref{fig:one} and~\ref{fig:two} show, as
long-dashed lines, the growth rates and wavenumber properties of
fastest-growing modes in our fiducial model for a high atmospheric
layer (model H1 in Table~2). Although strong similarities exist
between models H1 and B1 (solid line), especially in terms of
wavenumber properties, a striking difference is the absence of any
unstable mode in the vicinity of the magnetic ``pole,'' at $\theta
\sim 30$~deg. As we shall see, this results from the length-scale of
potentially unstable modes exceeding the maximum value allowed by our
search, $l_{\rm max} \sim H$ (the local scale height).

Figure~\ref{fig:four} shows estimates of the efficiency of momentum
transport in the spherical-radial direction, based on fastest-growing
linear mode properties in the high atmospheric layer, as a function of
polar angle $\theta$. The solid, dotted and dashed lines correspond to
models H1, H3 and H2, respectively. The horizontal dash-dotted line
indicates again the efficiency typically required to affect the
rotation profile over a local scale height in $1$~s ($\sim
H^2$~cm$^2$~s$^{-1}$).

Model H3 (dotted line) is voluntarily artificial, in that it allows
the length-scale of unstable modes to be as large as the neutron star
radius, $R_{\rm NS}$ (taken to be $10$~km), rather than the local
scale-height of the high-atmospheric layer, $H \sim 150$~cm. The
efficiency of momentum transport in this artificial model is strongly
peaked around the magnetic ``pole,'' at $\theta \sim 30$~deg, and much
resembles that of model B3 (solid line in
Fig.~\ref{fig:three}). Clearly then, the difference with model H1, and
in particular the absence of any unstable mode in the vicinity of the
magnetic ``pole,'' results from the tendency of these modes to have
large length-scales, in excess of $l_{\rm max} \sim H$, for typical
conditions high-up in the atmosphere.

This property is consistent with the gradual breakdown, as one goes up
in the atmosphere, of the important weak-field assumption made in our
analysis. Indeed, while the magnetic field strength may not vary
significantly over a few scale heights, density and pressure do. As a
result, our working hypothesis of a magnetic field weak compared to
thermal pressure becomes increasingly difficult to
satisfy. Specifically, while the Alfv\'en speed, $v_A \propto B
\rho^{-1/2}$, increases in the more tenuous high atmosphere, the sound
speed, $c_S \propto T^{1/2}$, decreases and the requirement $v_A \ll
c_S$ becomes less and less valid. It is only marginally satisfied in
the high-atmospheric layer we have considered ($v_A / c_S \sim 0.1$
for the values of $B_p$ adopted) and this is the origin of the
disappearance of unstable modes around the magnetic ``pole'' in model
H1 (e.g. Fig.~\ref{fig:four}). While large-scale displacements subject
to weak enough magnetic tension for instability could be found in
principle, they are clearly outside the regime of validity of our
weak-field analysis. Even higher up in the atmosphere, the weak field
regime breaks down completely and all unstable diffusive
magneto-rotational modes disappear.

Still, some of the fastest-growing modes identified in our realistic
high-atmosphere H1 model (solid line in Fig.~\ref{fig:four}) could
affect rotation profiles over a local scale height on rather short
time-scales, especially near the stellar equator.  Provided that these
unstable modes act to reduce the amount of differential rotation
initially present, a rotational recoupling of unstable layers in the
high atmosphere would then be expected on observationally interesting
time-scales.  It is important to notice, however, that the magnitude
of microscopic viscosity ($\nu + \nu_r$) in high-atmospheric layers
(compare Table~1 and Fig.~\ref{fig:four}) is sufficient by itself to
reduce the level of differential rotation on similar time-scales, so
that the existence of unstable modes, although important, is not
required to achieve rotational recoupling in these high layers.

Model H2 (dashed line in Fig.~\ref{fig:four}), differing from model H1
only by a smaller rate of spherical-radial differential rotation, $\dd
\ln \Omega / \dd \ln r = -1$, shows that our conclusions on the
stability and transport properties of diffusive magneto-rotational
modes do not strongly depend on the exact level of differential
rotation assumed to be present.

Finally, for completeness, let us mention that, by enforcing $B_p=0$,
we have been able to identify purely hydrodynamical multi-diffusive
(GSF) modes in high-atmospheric layers. These modes were characterized
by systematically lower momentum transport efficiencies than their
magnetized counterparts. They did not exist in lower atmospheric
layers, probably because local conditions are less double-diffusive as
one approaches the burning layers (see Table~1). This result is
consistent with the findings of Menou et al. (2004): a weak magnetic
field acts as a catalyst for rotational instability in a
multi-diffusive context, much like it does in an adiabatic context.

\subsection{Middle Atmosphere}
 
\begin{figure}
\psfig{figure=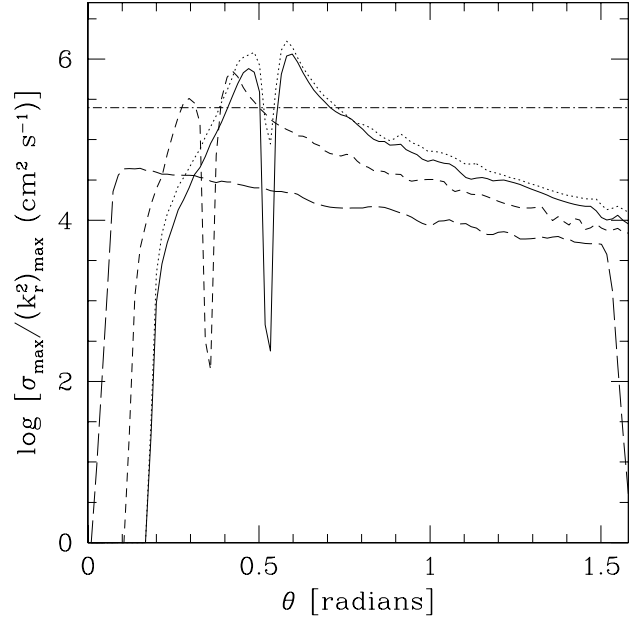,width=3.5truein,height=3.5truein}
\caption{Estimate of the efficiency of spherical-radial momentum
transport for typical conditions in a middle-atmospheric layer, as a
function of polar angle, $\theta$.  The solid, dotted, long-dashed and
short-dashed lines correspond to models M1, M4, M2 and M3 in Table~2,
respectively. The horizontal dash-dotted line indicates the efficiency
required to affect the rotation profile over a local scale-height in
$1$~s.}
\label{fig:five}
\end{figure}

\begin{figure}
\psfig{figure=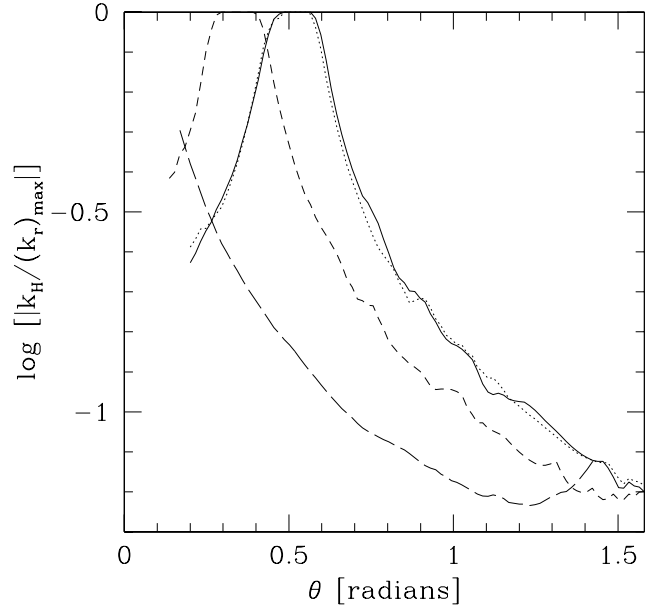,width=3.5truein,height=3.5truein}
\caption{Ratio of spherical-radial wavelength to local scale height,
noted $|k_{\rm H}/k_r|$, for the fastest-growing modes in a middle
atmospheric layer, as a function of polar angle, $\theta$. Notation is
identical to Fig.~5, for the same four models (M1: solid; M2:
long-dashed; M3: short-dashed; M4: dotted).}
\label{fig:six}
\end{figure}

Having identified important differences between unstable modes in
burning and high-atmospheric layers, it is only natural to wonder what
would the properties of these modes be in intermediate
middle-atmosphere conditions. Figure~\ref{fig:five} shows estimated
efficiencies of momentum transport, based on fastest-growing linear
mode properties, in four different models for middle-atmospheric
conditions. The solid, dotted, long-dashed and short-dashed lines
correspond to models M1, M4, M3 and M2 in Table~2, respectively. The
horizontal dash-dotted line shows the typical efficiency required to
affect rotation profiles over a local scale height in
$1$~s. Figure~\ref{fig:six} shows, for the same four models (and same
notation as in figure~~\ref{fig:five}), the ratio of the
spherical-radial wavelength of the fastest-growing mode to the local
scale height, noted $|k_H/k_r|$.

Results for middle-atmospheric layers shown in figures~\ref{fig:five}
and~\ref{fig:six} are indeed consistent with the conclusions reached
for burning and high-atmospheric layers. In all three models with a
misaligned dipole (M1, M3 and M4; Table~2), the effect of the magnetic
field geometry is readily seen. In the vicinity of the magnetic
``pole,'' fastest-growing diffusive magneto-rotational modes are
characterized by preferentially spherical-radial displacements (nearly
along field lines), with a radial wavelength reaching the maximum
value, $l_{\rm max} \sim H$ (Fig.~\ref{fig:six}), because the weak
magnetic field assumption is not ideally satisfied even for these
middle-atmospheric conditions. As a result, the modes left unstable in
the vicinity of the magnetic pole, which marginally satisfy the
constraint on $l_{\rm max}$, have slower growth rates and reduced
efficiencies of momentum transport. This is the origin of the dips in
figure~\ref{fig:five} and plateaus in figure~\ref{fig:six}, at $\theta
\sim \theta_{\rm mis}$. The model with an aligned dipole, on the other
hand, is characterized by fastest-growing modes with smaller radial
wavelengths and transport efficiencies.

What makes middle-atmospheric layers interesting is that, in the
models with a misaligned magnetic dipole, estimated efficiencies of
momentum transport are in excess of the reference value indicated by
the horizontal dash-dotted line, in the vicinity of $\theta \sim
\theta_{\rm mis}$. These estimates are also much in excess of the
transport caused by microscopic viscosity in these layers ($\nu_{\rm
tot} = \nu + \nu_r \sim 3 \times 10^3$~cm$^2$~s$^{-1}$;
Table~1). Consequently, the influence of these diffusive
magneto-rotational modes is expected to dominate the rotational
evolution of middle-atmospheric layers, on short enough timescales to
possibly have observational consequences. This conclusion is not very
sensitive to the exact value of the neutron star angular velocity,
$\Omega$. A comparison between models M1 and M4 (solid and dotted
lines in Fig.~\ref{fig:five}) shows that a faster stellar rotation
leads to faster growth and transport, but only moderately so.

\section{Discussion and Conclusion}

The atmospheres of weakly-magnetized neutron stars are subject to
vertical differential rotation during thermonuclear X-ray
bursts. Conditions in these atmospheres are also strongly
multi-diffusive and thus favor the development of magneto-rotational
instabilities. We have presented a linear stability analysis of the
expected differential rotation, with respect to local axisymmetric MHD
perturbations, in three representative atmospheric layers. Unstable
diffusive magneto-rotational modes were identified in much of the
differentially-rotating atmosphere. This indicates that momentum
transport between the various atmospheric layers could occur.

Our main motivation to perform a stability analysis is the possibility
that the transport resulting from instabilities may have observational
consequences for millisecond burst oscillations and associated
frequency drifts seen so far in a dozen accreting neutron star systems
(nuclear-powered X-ray pulsars; e.g. Strohmayer \& Bildsten 2003). The
differential rotation established during thermonuclear X-ray bursts is
transient, lasting perhaps several tens of seconds (a typical burst
duration). Therefore, it is not sufficient to identify
magneto-rotational instabilities in this context. For these
instabilities to have any effect, they must act to modify atmospheric
rotation profiles on short enough timescales. We have attempted to
address this issue by using linear estimates of the efficiency of
momentum transport and have concluded that, indeed, these
instabilities could be efficient enough to be of relevance to the bust
oscillation phenomenon.

There are reasons to believe that diffusive magneto-rotational
instabilities will, in a stellar context, tend to reduce the level of
existing differential rotation and perhaps make the system approach a
state of solid body rotation (see discussion in Menou et al. 2004; see
also Korycansky 1991). If correct, this conjecture would have
interesting consequences for the coherence of burst
oscillations. Indeed, it was noted by Cumming \& Bildsten (2000) that
the strong coherence of some burst oscillations implies, in the
absence of any momentum transport between atmospheric shells, that the
feature at the origin of the oscillations must be confined to a
surprisingly thin layer (much thinner than a local scale height in
their burning layer scenario). This feature would otherwise be smeared
out by differential rotation and the oscillations would lose their
coherence.

If, however, magneto-rotational instabilities act to reduce the amount
of differential rotation present during bursts, as we have suggested,
they would effectively promote the coherence of burst
oscillations. Constraints on the thickness of the layer from which the
modulated signal originates would then be relaxed. Magneto-rotational
instabilities would also contribute to the observed frequency drifts
of burst oscillations if they act efficiently enough to modify the
rotational evolution of atmospheric layers. This possibility is
consistent with the results of Cumming et al. (2002), who concluded
that conservation of specific angular momentum cannot account for the
largest observed frequency drifts.

As attractive as it is, this scenario suffers from a number of
significant uncertainties. For instance, we have cautioned about the
risks of estimating efficiencies of momentum transport from linear
mode properties only.  Also, the feature which is at the origin of
burst oscillations must be non-axisymmetric. The assumption of
axisymmetry made in our analysis could therefore be a limitation. We
have illustrated how the properties of fastest growing
magneto-rotational modes strongly depend on the local magnetic field
geometry. This may be viewed as a model uncertainty because the global
field topology on a given neutron star is a priori unknown. On the
other hand, this sensitivity to magnetic field geometry might in
principle be related to the link recently established between the
phases of persistent and burst oscillations, because it suggests a
special role for magnetic polar caps in the burst oscillation
phenomenon (Chakrabarty et al. 2003; Strohmayer et al. 2003). One then
wonders whether the diversity in drift timescales could be attributed
to different field geometries in different systems, perhaps caused by
varying degrees of field burying under continuous accretion (Cumming,
Zweibel \& Bildsten 2001).

Another possible limitation of our work comes from neglecting the
stabilizing role of composition gradients. In the vicinity of burning
layers, substantial composition gradients may be expected in ashes,
following the main episode of nuclear burning. The stratification due
to composition gradients is important because, contrary to thermal
stratification, it is not sensitive to a fast rate of heat
transfer. Although it is difficult to assess the importance of
stabilization due to composition gradients without a detailed study,
the specific example described by Goldreich \& Schubert (1967), for a
purely hydrodynamical situation, suggests that this effect should be
included when discussing neutron star atmosphere during
bursts. Another motivation to elucidate the role of stabilizing
composition gradients is the possibility that non-axisymmetric
versions of diffusive magneto-rotational modes could contribute to the
friction term entering the burning front models of Spitkovsky et
al. (2002).

Rather than focusing on burning layers only, we have performed
stability analyses at various heights in the differentially-rotating
neutron star atmosphere. One motivation for this is the stabilizing
effect of composition gradients that we have just discussed. Even if
such stabilization occurs in the vicinity of burning layers,
magneto-rotational instabilities may still be able to operate
higher-up in the atmosphere. A second motivation is the indication,
from the work of Cumming et al. (2002; see, e.g., their figure~4),
that layers above the burning regions are involved in the burst
oscillation phenomenon, since they are the only ones having the right
amount of rotational offset to explain the largest observed frequency
drifts.

Our analysis for high atmospheric layers shows that magneto-rotational
instabilities cease gradually to exist as one approaches the stellar
photosphere, because the weak magnetic tension they require becomes
increasingly difficult to satisfy. Still, according to our
middle-atmospheric models, there is a region above the burning layers
in which magneto-rotational instabilities can operate and possibly
lead to a rather efficient transport of momentum. While burst
oscillations have traditionally been associated with burning layers,
it may be worth giving more attention to these middle-atmospheric
layers. Not only do they have the right amount of rotational offset to
explain large frequency drifts, but these layers are also more
advantageous for the visibility of oscillations, with a short heat
diffusion time compared to the time they take to revolve around the
star (see Cumming \& Bildsten 2000 for a discussion of the visibility
of burst oscillations).

We note that Cumming \& Bildsten (2000) and Cumming et al. (2002) have
invoked the mechanism of magnetic field wind-up described by Spruit
(1999) as a possible source of recoupling between
differentially-rotating layers. In regions with weak enough
stratification from composition gradients for magneto-rotational
instabilities to operate, this process may not be relevant. Indeed, in
the case of differentially-rotating accretion disks, it has been
argued analytically and demonstrated numerically that the linear
growth of the azimuthal field component caused by differential
rotation is not important because of the exponential growth of
magneto-rotational modes (Balbus \& Hawley 1991; 1998). By analogy,
one may expect the same linear growth to be irrelevant in the stellar
context, given the existence of exponentially-growing diffusive
magneto-rotational modes.

Beyond simple qualitative points, it is difficult to deepen an
interpretation of burst oscillations and associated frequency drifts
in terms of a rotational evolution which is partly driven by
magneto-rotational instabilities.  A time-dependent model for the
coupled thermal, chemical and rotational evolution of a neutron star
atmosphere during burst seems to be required to make more definite
predictions. Such time-dependent models, ignoring the rotational
evolution, already exist and exhibit complex behaviors. For example,
transient convection early in the burst leads to some mixing of
momentum and elements, while inverted composition gradients (which
would presumably promote instabilities) are sometimes obtained after
the main nuclear burning phase (see, e.g., Woosley \& Weaver 1984;
Woosley et al. 2003). In order to include the rotational evolution,
these models would require reliable estimates of the efficiency of
momentum transport (and of the resulting chemical mixing), which could
presumably be obtained from fully turbulent numerical simulations of
the non-linear development of diffusive magneto-rotational
instabilities.  Until such advanced tools are developed, it is likely
that observations will remain our best guide for understanding burst
oscillations and associated frequency drifts.

\section*{Acknowledgments}

It is a pleasure to thank James Cho and Andrew Cumming for useful
discussions, Steve Balbus and Lars Bildsten for comments on an early
version of the manuscript, and the Department of Astronomy at the
University of Virginia for their hospitality.

\end{document}